\documentclass[12pt]{article}

\usepackage{amsmath}
\usepackage{amsfonts}
\usepackage{amssymb}
\usepackage{graphicx}
\usepackage{subfigure}
\usepackage{rotating}
\usepackage{color}
\usepackage{natbib, algorithm2e}
\usepackage{graphicx,setspace,amsmath,amssymb,subfigure,url,multirow,booktabs,
verbatim,bm, amsthm}

\allowdisplaybreaks[3]

\newtheorem{theorem}{Theorem}

\def\bse{\begin{eqnarray*}}
\def\ese{\end{eqnarray*}}
\def\beq{\begin{equation}}
\def\eeq{\end{equation}}
\def\bqe{\begin{eqnarray}}
\def\eqe{\end{eqnarray}}

\begin{document}


\thispagestyle{empty}

\title{\bf Robust Identification of Gene-Environment Interactions under High-Dimensional Accelerated Failure Time Models}

\author{Qingzhao Zhang$^{1\dagger}$,
Hao Chai$^{2\dagger}$, Shuangge Ma$^{2*}$\\
$^1$School of Economics and the Wang Yanan Institute for Studies in Economics,\\
Xiamen University\\
$^2$Department of Biostatistics, Yale University\\
$^\dagger$ {\it Equal contributions}\\
{\it *email: shuangge.ma@yale.edu}}
\date{}

\maketitle


\begin{abstract}
For complex diseases, beyond the main effects of genetic (G) and environmental (E) factors, gene-environment (G-E) interactions also play an important role. Many of the existing G-E interaction methods conduct marginal analysis, which may not appropriately describe disease biology. Joint analysis methods have been developed, with most of the existing loss functions constructed based on likelihood. In practice, data contamination is not uncommon.
Development of robust methods for interaction analysis that can accommodate data contamination is very limited.  In this study, we consider censored survival data and adopt an accelerated failure time (AFT) model. An exponential squared loss is adopted to achieve robustness. A sparse group penalization approach, which respects the ``main effects, interactions'' hierarchy, is adopted for estimation and identification. Consistency properties are rigorously established. Simulation shows that the proposed method outperforms direct competitors. In data analysis, the proposed method makes biologically sensible findings.
\end{abstract}

\noindent{\bf Keywords}: Gene-environment interaction; Robustness; Prognosis; Joint analysis.

\section{Introduction}\label{s:intro}
For many complex diseases, it is essential to identify important risk factors that are associated with prognosis. In the omics era, profiling studies have been extensively conducted. It has been found that, beyond the main effects of genetic (G) and environmental (E) risk factors, gene-environment (G-E) interactions can also have important implications.

Denote $T$ and $C$ as the prognosis and censoring times, respectively. Denote $X = (X_1, \ldots, X_q)^\top$ as the $q$ environmental/clinical variables, and $Z=(Z_1, \ldots, Z_p)^\top$ as the $p$ genetic variables. The existing G-E interaction analysis methods mainly belong to two families. The first family conducts marginal analysis \citep{hunter2005gene, Thomas2010methods, shi2014gene}, under which one or a small number of genes are analyzed at a time.
Despite its significant computational simplicity, marginal analysis contradicts the fact that the prognosis of complex diseases is attributable to the {\it joint} effects of multiple main effects and interactions. The second family of methods, which is biologically more sensible, conduct joint analysis \citep{liu2013identification, SIM:SIM6287, zhu2014identifying}. Among the existing joint analyses, the regression-based is the most popular and proceeds as follows. Consider the model $T \sim \phi( \alpha_{0}+\sum_{j=1}^qX_j\alpha_{j} + \sum_{k =1}^pZ_k\beta_k + \sum_{j = 1}^q\sum_{k=1}^p X_jZ_k\gamma_{j,k})$, where model $\phi(\cdot)$ is known up to the regression coefficients $\alpha_{0},\{\alpha_{j}\}_1^q, \{\beta_{k}\}_1^p$, and $\{\gamma_{j,k}\}_1^{q,p}$. Conclusions on the importance of interactions are drawn based on $\{\gamma_{j,k}\}_1^{q,p}$. With the high data dimensionality and demand for the selection of relevant effects, regularized estimation is usually needed.

In the dominating majority of the existing studies, estimation is based on the standard likelihood, which is nonrobust. In practice, data contamination is not uncommon and can be caused by multiple reasons.
Many diseases are heterogeneous, and different subtypes behave differently. When the subtype information is accurately available, subtype-specific analysis can be conducted. However, when such information is not or partially available, which is often the case in practice \citep{he2015robust}, subjects belonging to small subtypes may be viewed as ``contamination'' to those of the leading subtype.
Human errors can also happen. It has been well noted that survival information extracted from medical records is not always reliable
\citep{fall2008reliability, bowman2011doctors}, creating contamination in prognosis distributions. In low-dimensional biomedical studies, it has been well established that even a single contaminated observation can lead to biased model estimation and so false marker identification \citep{huber2009robust}. Our literature review suggests that in the analysis of G-E interactions, robust methods that can effectively accommodate contamination in prognosis outcomes have been very rare. For marginal interaction analysis, a few robust methods, for example the multifactor dimensionality reduction (MDR), have been developed. However, they are not directly applicable to joint analysis because of both methodological and computational challenges. As discussed in \cite{wu2014selective}, a handful of robustness studies have been conducted under high-dimensional settings for joint analysis. However, they are mostly on main effects and not directly applicable to interaction analysis because of the additional complexity caused by the ``main effects, interactions'' hierarchy. Most of them adopt the quantile regression technique. Studies under low-dimensional settings suggest that no robust technique can dominate. It is thus desirable to examine alternative robust techniques under high-dimensional settings. In addition, for quite a few existing methods, statistical properties have not been well studied, casting doubts on their validity.

Consider data with a prognosis outcome and both G and E measurements. Our goal is to conduct joint analysis and identify important G-E interactions and main G and E effects. This study advances from the literature in multiple aspects. Specifically, we consider the scenario with possible contamination in the prognosis outcome, which is commonly encountered but little addressed. We adopt an exponential squared loss to achieve robustness. This loss function provides a useful alternative to the popular quantile regression and other robust approaches but has not been well investigated under high-dimensional settings, especially not for interaction analysis. This study also marks a novel extension of the exponential squared loss to accommodate censored survival data. For regularized estimation and selection of relevant effects, we propose adopting a penalization technique, which respects the ``main effects, interactions'' hierarchy. Significantly advancing from most of the existing studies, consistency properties are rigorously established. Theoretical research for high-dimensional robust methods remains limited. As such, this study may provide valuable insights beyond this article. With both methodological and theoretical developments, this study is warranted beyond the existing literature.

\section{Methods}
\label{s:interaction}

\subsection{Data and model settings}

For describing prognosis, we adopt the AFT model, which has been the choice of multiple studies with high-dimensional genetic data \citep{liu2013identification, shi2014gene}. Compared to alternatives including the Cox model, advantages of the AFT model include intuitive interpretations and low computational cost, which are especially desirable with high-dimensional genetic data. With a slight abuse of notation, still use $T$ and $C$ to denote the logarithms of the event and censoring times, and $\delta = I_{\{T\leq C\}}$. The AFT model specifies that
$$T =\alpha_0 +\sum_{j=1}^qX_j \alpha_j+ \sum_{k=1}^pZ_k\beta_k + \sum_{j=1}^{q}\sum_{k=1}^p X_jZ_k\gamma_{j,k}+\epsilon,$$
where $\epsilon$ is the random error. Following \cite{stute1993consistent,stute1996distributional}, we assume that $T$ and $C$ are independent, and $\delta$ is conditionally independent of $(X^\top, Z^\top)^\top$ given $T$. Let $W_k= (Z_k, X_1Z_k, \ldots, X_q Z_k)^\top$ and $b_k=(\beta_k, \gamma_{1,k}, \ldots, \gamma_{q,k})^\top$, which represent all main and interaction effects corresponding to the $k$th genetic variable.

With $n$ independent subjects, use subscript ``$i$" to denote the $i$th subject. For subject $i$, let $y_i = \min\{T_i, C_i\}$ and $\delta_i=I_{\{T_i\leq C_i\}}$ be the observed time and event indicator, respectively. Then the $i$th observation consists of $(y_i, \delta_i, \mathbf{x}_i, \mathbf{z}_i)$, with $\mathbf{x}_i=(x_{i1},\ldots, x_{iq})^\top$, $\mathbf{z}_i=(z_{i1},\ldots, z_{ip})^\top$, and $W_{k,i} = (z_{ik}, x_{i1}z_{ik}, \ldots, x_{iq}z_{ik})^\top$ denoting the $i$th realization of $X$, $Z$, and $W_k$, respectively. Denote $\mathbf{u}_{i,}^\top=(1,\mathbf{x}_i^\top,W_{1,i}^\top, \ldots, W_{p,i}^\top)$, $\mathbf{U}= (\mathbf{u}_{1,}, \cdots, \mathbf{u}_{n,})^\top$, and  $\bm\zeta=(\alpha_0,\ldots,\alpha_q, b_1^\top,\ldots,b_p^\top)^\top$. Without loss of generality, assume that $(y_i,\delta_i,\mathbf{u}_{i,})$'s have been sorted according to $y_i$'s in an ascending manner.

\subsection{Robust estimation and identification}
Consider the scenario where the distribution of $\epsilon$ is not specified, which significantly differs from the existing parametric studies and makes the proposed method more flexible. To motivate the proposed estimation, first consider data without contamination. \cite{stute1993consistent} developed a weighted least squared estimation approach. Under low-dimensional settings, Stute's estimator is defined as the minimizer of the loss function
\begin{equation*}
  \sum_{i=1}^n\omega_i(y_i-\mathbf{u}_{i,}^\top{\bm\zeta})^2.
\end{equation*}
Here the weights $\bm\omega=(\omega_i)_{i=1}^n$ are computed based on the Kaplan-Meier estimation and defined as
$$\omega_{1}=\frac{\delta_{1}}{n},~\omega_{i}=
\frac{\delta_{i}}{n-i+1}\prod_{j=1}^{i-1}
(\frac{n-j}{n-j+1})^{\delta_{j}},~i=2,\cdots,n.$$
It is noted that Stute's estimator is not necessarily the most efficient. However, under high-dimensional settings, it can be computationally the most convenient with the least squared loss. It can be seen that, if $\omega_i \neq 0$, one contaminated $y_i$ can lead to severely biased model estimation.

Now consider the scenario with possible outliers in the prognosis data. We propose the objective function
\begin{eqnarray}
	Q_{\theta}(\bm\zeta)=
	\sum_{i=1}^n\omega_i\exp(-(y_i - \mathbf{u}_{i,}^\top\bm\zeta)^2 / \theta).
	\label{eqn1}
\end{eqnarray}
This function has been motivated by the following considerations. Under low-dimensional regression analysis without censoring, \cite{wang2013robust} adopted an exponential squared loss to achieve robustness. The intuition is as follows. For a contaminated subject with the observed $y_i$ deviating from $\mathbf{u}_{i,}^\top\bm\zeta $ (the ``predicted" value based on the model), $(y_i - \mathbf{u}_{i,}^\top\bm\zeta)^2$ has a large value. The exponential function down-weighs such a contaminated observation. The degree of down-weighing is adjusted by $\theta$: when $\theta$ gets smaller, the contaminated observations have smaller influence. While sharing certain similar ground as \cite{wang2013robust} and others, the present study has three main challenges/advancements. The first is the high dimensionality, which brings tremendous challenges to theoretical and computational developments. The second is the need to respect the ``main effects, interactions'' hierarchy (more details below). The third is censoring, to accommodate which we introduce the weight function $\omega_i$ motivated by Stute's approach. As the weights are data-dependent, they bring challenges to the establishment of theoretical properties.

When $p\gg n$, regularized estimation is needed. In addition, out of a large number of profiled G factors and G-E interactions, only a few are expected to be associated with prognosis. We adopt penalization for regularized estimation and identification, which has been the choice of a large number of genetic studies, especially recent interaction analyses \citep{bien2013lasso, liu2013identification, shi2014gene}. Specifically, consider the penalized robust objective function
\begin{equation}
	L_{\lambda_1, \lambda_2, \theta}(\bm\zeta)=Q_{\theta}(\bm\zeta)-\sum_{k=1}^{p}\rho(\|b_k\|; \lambda_1, s)-\sum_{k=1}^{p}\sum_{j=2}^{q+1}\rho(|b_{kj}|; \lambda_2, s),
	\label{eqn9}
\end{equation}
where $\|\cdot\|$ is the $\ell_2$ norm, $\rho(t; \lambda, s) = \lambda \int_0^{|t|} \left(1-\frac{x}{\lambda s} \right)_+ dx$ is the MCP(minimax concave penalty, \cite{zhang2010nearly}), and $b_{kj}$ is the $j$th element of $b_{k}$. $\lambda_1$ and $\lambda_2$ are data-dependent tuning parameters, and $s$ is the regularization parameter,  per the terminologies in \cite{zhang2010nearly}. The robust estimator is defined as the maximizer of $L_{\lambda_1, \lambda_2, \theta}(\bm\zeta)$. An interaction term (or main effect) is concluded as important if its estimate is nonzero.

In recent genetic interaction analysis, it has been stressed that the ``main effects, interactions'' hierarchy should be respected. That is, if an interaction term is identified as important, its corresponding main effect(s) should be automatically identified. G-E interaction analysis has its uniqueness. The E variables usually have a low dimensionality and are manually chosen. As such, selection is usually not conducted on the E variables (if desirable, this can be easily achieved). Thus for G-E interaction analysis, the hierarchy postulates that if an G-E interaction is identified as important, its corresponding main G effect is automatically identified.
In the adopted sparse group penalty, the first penalty, which is a group MCP, determines which groups are selected. Here one group corresponds to one genetic variable and its interactions. As the group MCP does not have within-group sparsity, the second penalty is imposed, where we penalize the interaction terms and determine which are nonzero. With the special design that the second penalty is only imposed on interactions, important interactions correspond to important groups, automatically leading the estimates of the corresponding main G effects nonzero. As such, the combination of the two penalties guarantees the hierarchy.
We note that although sparse group penalization has been studied in the literature \citep{liu2013identification}, it has been very rarely coupled with robust loss functions. It is also noted that MCP can be potentially replaced by other penalties.

\subsection{Computation}

In this section, we develop an efficient algorithm to compute the maximizer of $L_{\lambda_1, \lambda_2, \theta}(\bm\zeta)$. The basic strategy is to iteratively approximate the objective function by its quadratic minorization. Then a coordinate-wise updating procedure is used to find the maximizer of each approximated objective function. The maximizer then serves as the starting point for the next minorization. Overall, this is a coordinate-descent (CD) algorithm nested in a Minorize-Maximization (MM) algorithm.

Let $\mathbf{W}(\bm\zeta)$ be a diagonal matrix with the $i$th diagonal element $\mathbf{W}_{i,i}=2\omega_i\exp(-(y_i-\mathbf{u}_{i,}^\top\bm\zeta)^2/\theta)/\theta$. Also let $\mathbf{v}(\bm\zeta) = (v_1, \cdots, v_n)^\top$ with $v_i=y_i-\mathbf{u}^\top_{i,}\bm\zeta$. Define $\mathbf{U}_{,-j}$ as the sub-matrix of $\mathbf{U}$ with the $j$th column excluded.  Define $\mathbf{u}_{,j}$ as the $j$th column of matrix $\mathbf{U}$, and $u_{i,j}$ as the $j$th component of vector $\mathbf{u}_{i,}$. Similarly, define $\bm{\zeta}_{-j}$ as the sub-vector of $\bm\zeta$ with the $j$th element excluded. For the exponential squared objective function in~\eqref{eqn1}, its first and second order derivatives with respect to $\bm\zeta$ are
\begin{eqnarray*}
	\frac{\partial{Q_\theta(\bm\zeta})}{\partial \zeta_j} &=& 2\sum_{i=1}^n\omega_i\exp(-(y_i - \mathbf{u}_{i,}^\top\bm\zeta)^2 / \theta)u_{i,j}(y_i - \mathbf{u}_{i,}^\top\bm\zeta)/\theta=\mathbf{u}_{,j}^\top \mathbf{W}(\bm\zeta)\mathbf{v}(\bm\zeta),\\
	 \frac{\partial^2{Q_\theta(\bm\zeta)}}{\partial\zeta_{j}\partial\zeta_k}&=&
2\sum_{i=1}^n\omega_i\exp(-(y_i - \mathbf{u}_{i,}^\top\bm\zeta)^2 / \theta)u_{i,j}u_{i,k}[2(y_i - \mathbf{u}_{i,}^\top\bm\zeta) ^ 2/\theta-1]/\theta.
\end{eqnarray*}
If $(y_i-\mathbf{u}^\top_{i,}\bm\zeta) ^ 2 /\theta >0.5$, $\frac{\partial^2{Q_\theta(\bm\zeta)}}{\partial\zeta_{j}\partial\zeta_k}\geq 0$. On the other hand if $(y_i-\mathbf{u}^\top_{i,}\bm\zeta)^2 /\theta \leq 0.5$, $\frac{\partial^2{Q_\theta(\bm\zeta)}}{\partial\zeta_{j}\partial\zeta_k}\leq 0$. Hence, to find the maximizer of $Q_\theta(\bm\zeta)$, the simple Newton-Raphson approach may lead to infinity if the starting value is too far from the true value. To tackle this problem, a minorization of $Q_\theta(\bm\zeta)$ is used to approximate $Q_\theta(\bm\zeta)$. Note that $\frac{\partial^2{Q_\theta(\bm\zeta)}}{\partial\zeta_{j}\partial\zeta_k}\geq-
2\sum_{i=1}^n\omega_i\exp(-(y_i - \mathbf{u}_{i,}^\top\bm\zeta)^2 / \theta)u_{i,j}u_{i,k}/\theta$. Hence a minorized approximation to $Q_\theta(\bm\zeta)$ at $\bm\zeta^{m}$ is
\begin{equation*}
	Q_\theta(\bm\zeta^m) + \mathbf{v}^\top(\bm\zeta^m)\mathbf{W}(\bm\zeta^m)\mathbf{U}(\bm\zeta-\bm\zeta^m)
-\frac12(\bm\zeta-\bm\zeta^m)^\top\mathbf{U}^\top \mathbf{W}(\bm\zeta^m)\mathbf{U}(\bm\zeta-\bm\zeta^m).
\end{equation*}
Note that $\bm\zeta^m=(\alpha_0^m,\ldots,\alpha_q^m, {b_1^m}^\top,\ldots,{b_p^m}^\top)^\top$ with $b_k^m=(\beta_k^m, \gamma_{1,k}^m, \ldots, \gamma_{q,k}^m)^\top$. For the penalty, we apply a local linear approximation at $\bm\zeta^{m}$, which is given by
\[-\sum_{k=1}^p\dot{\rho}(\|b_k^m\|;\lambda_1, s)\frac{|\beta_k^m|}{\|b_k^m\|}|\beta_k| - \sum_{k=1}^p\sum_{j=1}^q \left\{\dot{\rho}(\|b_k^m\|;\lambda_1, s)\frac{|\gamma_{j,k}^m|}{\|b_k^m\|}+ \dot{\rho}(|\gamma_{j,k}^m|;\lambda_2, s)\right\}|\gamma_{j,k}|  \]
if the terms that do not depend on $\bm\zeta$ are ignored, where $\dot{\rho}(t;\lambda, s)  =    \text{sgn}(t) \left(\lambda - \frac{|t|}{s}\right)_{+}.$
If we replace $Q_\theta(\bm\zeta)$ in~\eqref{eqn9} with its minorized approximation and plug in the approximation of the penalty, the penalized objective function then has the form
\begin{eqnarray}\label{eqn17}
&&	L_{\lambda_1, \lambda_2,\theta}(\bm\zeta|\bm\zeta^m) = Q(\bm\zeta^m) + \mathbf{v}^\top(\bm\zeta^m)\mathbf{W}(\bm\zeta^m)\mathbf{U}
(\bm\zeta-\bm\zeta^m)-
\frac12(\bm\zeta-\bm\zeta^m)^\top\mathbf{U}^\top \mathbf{W}(\bm\zeta^m)\mathbf{U}(\bm\zeta-\bm\zeta^m)\nonumber\\
&&~-\sum_{k=1}^p\dot{\rho}(\|b_k^m\|;\lambda_1, s)\frac{|\beta_k^m|}{\|b_k^m\|}|\beta_k| - \sum_{k=1}^p\sum_{j=1}^q \left\{\dot{\rho}(\|b_k^m\|;\lambda_1, s)\frac{|\gamma_{j,k}^m|}{\|b_k^m\|}+ \dot{\rho}(|\gamma_{j,k}^m|;\lambda_2, s)\right\}|\gamma_{j,k}| .
\end{eqnarray}
This function has a ``weighted quadratic + penalty'' form and can be optimized using the coordinate descent approach.

The algorithm starts with $m=0$ and $\bm\zeta^m = \mathbf{0}$, where $m$ is the index of the MM iteration. At iteration $m$, the objective function is approximated by its minorization $L_{\lambda_1, \lambda_2, \theta}(\bm\zeta|\bm\zeta^m)$ given in~\eqref{eqn17}. Then the penalized weighted quadratic function is maximized using the coordinate descent algorithm. Denote $\bar{\bm\zeta}^{old}$ as the estimate of $\bm\zeta$ before updating. We update each element of the estimate, and denote the new estimate as $\bar{\bm\zeta}^{new}$. This is repeated until the distance between $\bar{\bm\zeta}^{old}$ and $\bar{\bm\zeta}^{new}$ is smaller than a prefixed constant. Then $\bm\zeta^{m+1} = \bar{\bm\zeta}^{new}$ serves as the new expansion base point for the next minorization. The overall procedure is repeated until convergence. Convergence properties of the MM and CD techniques have been well studied in the literature. With our problem, the objective function increases at each step and is bounded above, which leads to convergence. In numerical study, we conclude convergence if the difference between two estimates after two consecutive MM steps is small enough. We observe convergence in all numerical examples after a small to moderate number of MM iterations.

The proposed method involves  tuning parameters. For $s$ in MCP, we follow \cite{zhang2010nearly} and other published studies, which suggest examining a small number of values or fixing it. In our numerical study, we fix $s=6$, which has been adopted in published studies \citep{shi2014gene, xu2018robust}. We have also examined $s$ values near 6 and observed similar performance (details omitted). In practice, for settings significantly different from ours, other $s$ values may need to be considered. Under low-dimensional settings, \cite{wang2013robust} proposed an iterative approach to select the robust tuning parameter $\theta$. However, their approach is computationally infeasible for high-dimensional data. Under the present setting, for each combination of $(\lambda_1, \lambda_2, \theta)$,  we compute the solution. This way, we can obtain a solution surface over a three-dimensional tuning parameter grid. This is feasible as the proposed computational algorithm only involves simple updates and incurs low cost. Then the tuning parameters can be selected using a prediction-based method which proceeds as follows: (a) compute the cross-validated sum of prediction errors for each $(\lambda_1, \lambda_2,\theta)$ combination; (b) for each fixed $\theta$, average the sum of prediction errors over $\lambda_1, \lambda_2$. Select $\theta$ that has the smallest average sum of prediction errors; (c) with the selected $\theta$, select $\lambda_1, \lambda_2$ that has the smallest sum of prediction errors. This procedure first groups all $(\lambda_1, \lambda_2)$ values together and selects the best $\theta$ value. Then with the optimal $\theta$ value, the optimal $(\lambda_1, \lambda_2)$ values are selected. Our numerical experiments suggest that this procedure generates more stable estimates than directly searching over the three-dimensional $(\lambda_1, \lambda_2, \theta)$ grid.

With a complex robust goodness-of-fit and a penalty that respects the hierarchy, the proposed method is inevitably computationally more expensive than some simpler alternatives. However, as the proposed computational algorithm is composed of relatively simple calculations, the overall computational cost is affordable. With fixed tunings, the analysis of one simulated dataset (described in detail below) takes about nine minutes on a regular laptop. Tuning parameter selection can be conducted in a highly parallel manner to save computer time.

\subsection{Consistency properties}

In this section, we rigorously prove that the proposed method can consistently identify the important interactions (and main effects) under ultrahigh-dimensional settings. In the literature, theoretical development for robust methods under high-dimensional settings has been limited. It is especially rare for methods other than the quantile-based. With the consistency properties, the proposed method can be preferred over the alternatives whose statistical properties have not been well established. Our theoretical development not only provides a solid ground for the proposed method but also sheds insights for other robust methods under high-dimensional settings.

For any two subsets $S_1$ and $S_2$ of $\{1, \cdots, p+q+ pq+ 1\}$ and a matrix $H$, we denote by $H_{S_1 S_2}$ the submatrix of $H$ with rows and columns indexed by $S_1$ and $S_2$, respectively. Let $\bm\zeta^*=(\alpha_0^*,\ldots,\alpha_q^*, {b_1^*}^\top,\ldots,{b_p^*}^\top)^\top$, where $b_k^*=(\beta_k^*, \gamma_{1,k}^*, \ldots, \gamma_{q,k}^*)^\top$ is the true value of $\bm\zeta$. Here we make the sparsity assumption, under which  only a subset of the components of $\bm\zeta^*$ is nonzero. Define the three groups of parameters:
\begin{eqnarray*}
&&A_1 = \{\alpha_0^*, \ldots, \alpha_q^*\}, ~~~A_2 = \{\gamma_{j,k}^*: \gamma_{j,k}^*\neq 0,  j=1,\ldots, q; k=1, \ldots, p\}, \\
&&A_3 = \{\beta_k^*: \beta_k^*\neq 0~\mbox{or there exsits some }~1\leq j\leq q~\mbox{such that}~\gamma_{j,k}^*\neq 0, k=1, \ldots, p \}.
\end{eqnarray*}
Denote $\mathcal{A}$ as the set of indices of $A_1\cup A_2 \cup A_3$ in the vector ${\bm\zeta}^*$.  Let $\mathcal{A}^c$ and $|\mathcal{A}|$ denote the complement and cardinality of set $\mathcal{A}$, respectively. We then divide $\mathcal{A}^c$ into there sets of indices $\mathcal{B}_1, \mathcal{B}_2$, and $\mathcal{B}_3$, which correspond to the following three sets
\begin{eqnarray*}
	&&B_1 = \{\beta_k^*: \beta_k^*= 0, k=1, \ldots, p\}, ~B_2 = \{\gamma_{j,k}^*: \gamma_{j,k}^*= 0~\mbox{but}~\beta_k^*\neq 0,  j=1,\ldots, q; k=1, \ldots, p \}, \\
	&&B_3 = \{\gamma_{j,k}^*: \gamma_{j,k}^*= 0 ~\mbox{and}~\beta_k^*= 0 ,  j=1,\ldots, q; k=1, \ldots, p\},
\end{eqnarray*}
respectively.
Define $$D_{n}({\bm\zeta}) = \sum_{i=1}^n \omega_i \exp(-(y_i - \mathbf{u}_{i}^\top{\bm\zeta})^2 / \theta)\frac{2(y_i - \mathbf{u}_{i,}^\top{\bm\zeta})}{\theta}\mathbf{u}_{i,}$$
and
$$I_{n}({\bm\zeta}) = \frac{2}{\theta}\sum_{i=1}^n \omega_i \exp(-(y_i - \mathbf{u}_{i,}^\top{\bm\zeta})^2 / \theta)\left(\frac{2(y_i - \mathbf{u}_{i,}^\top{\bm\zeta})^2}{\theta}-1\right)\mathbf{u}_{i,}
\mathbf{u}_{i,}^\top .$$
The following conditions are needed to establish the consistency properties.

\begin{description}
\item[C1.] $T$ and $C$ are independent, and $P(T \leq C|T,X,Z) = P(T \leq C|T)$.
\item[C2.] The support of $T$ is dominated by that of $C$. For example, $\tau_T<\tau_C$ or $\tau_T= \tau_C=\infty$, where $\tau_T$ and $\tau_C$ are the right end points of the support of $T$ and $C$, respectively.
\item[C3.] $E[D_{n}({\bm\zeta}^*)] = 0$.
\item[C4.] The distributions of $D_{n,j}({\bm\zeta}^*)$'s are subgaussian, that is, $\Pr(|D_{n,j}({\bm\zeta}^*)| >t )\leq 2 \exp\left(- n t^2/\sigma^2\right)$. Moreover, $I_{n, jk}({\bm\zeta})- I_{jk}({\bm\zeta})$'s  are subgaussian for all $\bm\zeta\in \Theta =\{\bm\zeta: \|\bm\zeta- \bm\zeta^*\|_2 <\delta\}$, where $\delta$ is a positive constant,
$I({\bm\zeta}) = E[I_{n}({\bm\zeta})]$, and $I_{jk}({\bm\zeta})$ is the $(j,k)$th component of matrix $I({\bm\zeta})$. Moreover, there exists a bounded constant $\kappa$ such that ${\bm \nu}^\top [I({\bm\zeta}^1)-I({\bm\zeta}^2)]{\bm\nu} \leq \kappa \|{\bm\zeta}^1-{\bm\zeta}^2\|_2$ for any ${\bm\zeta}^1, {\bm\zeta}^2\in \Theta$ and $\|{\bm\nu}\|_2 = 1$.
\item[C5.] $I_{\mathcal{A}\mathcal{A}}({\bm\zeta}^*)$ is a $|\mathcal{A}|\times |\mathcal{A}|$ negative-definite matrix. The eigenvalues of $I_{\mathcal{A}\mathcal{A}}({\bm\zeta}^*)$ are bounded away from zero and infinity.
\item[C6.] $\min_{j,k} \{|\gamma_{j,k}^*|: \gamma_{j,k}^*\neq 0\}\gg \lambda_1\vee \lambda_2$.
$\lambda_1\wedge \lambda_2\gg \sqrt{|\mathcal{A}|/n}$.
\end{description}
C1 and C2 have been commonly assumed in the literature. See for example \cite{stute1993consistent},~\cite{stute1996distributional}, and \cite{huang2007least}. We note that the independent censoring assumption usually holds in practice, although from a theoretical perspective, quite a few studies have made the weaker conditional independence assumption. We have explored relaxing this assumption and found that alternative and less intuitive assumptions would have to be made. The zero expectations in C3 and C5 ensure the consistency of estimation. C4 is required for Theorem 1, and a similar assumption has been made in \cite{ma2012variable}.  C6 requires that the smallest signal does not decay too fast, which is common in studies on high-dimensional inference. The following theorem establishes consistency of the proposed estimator $\widehat{\bm\zeta}$.

\begin{theorem}\label{11}
 Suppose that conditions C1-C6 hold. Let $\varpi_n= (\lambda_1\wedge \lambda_2)/\{\max(\Phi_1, \Phi_2, \Phi_3)\}$, where $\Phi_t= \|I_{\mathcal{B}_t \mathcal{A}}({\bm\zeta}^*)I_{ {\mathcal{A}\mathcal{A}}}({{\bm\zeta}}^*)^{-1}\|_\infty$, $t=1,2,3$. If $|\mathcal{A}| =o(n)$, $\lambda_1\vee \lambda_2 \rightarrow 0$, $n \varpi_n^2 \rightarrow \infty$,  and $\log p = o ( n\varpi_n^2)$, with probability tending to one, we have
\begin{equation*}(a)~~~\|\widehat{\bm\zeta}_\mathcal{A} - {\bm\zeta}_\mathcal{A}^*\|_2= O_p(\sqrt{|\mathcal{A}|/n});~~~~ (b)~~~ \widehat{\bm\zeta}_{\mathcal{A}^c}= \textbf{0}.
\end{equation*}
\end{theorem}
This theorem establishes that the proposed method is able to accommodate $p$ with $\log p = o ( n\varpi_n^2)$. The penalized robust estimator enjoys the same asymptotic properties as the oracle estimator with probability approaching one. This property holds under high dimensions without restrictive conditions on the errors. To the best of our knowledge, properties of the robust exponential loss, even without censoring, have not been studied under high-dimensional settings. Thus our theoretical investigation can have independent value. Proof of the theorem is presented in Appendix.

\section{Simulation}
\label{s:simu}
In simulation, we set $n = 300, q = 5$, and $p = 1000$. The underlying true model contains a total of 35 nonzero effects, including 5 main E effects, 10 main G effects, and 20 interactions. The ``positions'' of nonzero main G effects are randomly placed. The nonzero interactions are generated to respect the ``main effects, interactions'' hierarchy. The nonzero regression coefficients are randomly generated from $uniform(0.7,1.3)$. We consider both continuous and categorical distributions to mimic, for example, gene expression and SNP data. Specifically, under the continuous scenario, the E and G factors are generated from multivariate normal distributions with marginal means zero, marginal variances one, and the following variance matrix structures: Independent, AR(0.3), AR(0.8), Band(0.3), Band(0.6), and CS(0.2). Under the independent scenario, all factors have zero correlations. Under the AR$(\rho)$ correlation structure, for the $i$th and $j$th factors, $corr = \rho^{|i-j|}$. Under the Band$(\rho)$ correlation structure, for the $i$th and $j$th factors, $corr =\rho\cdot I(|i - j| = 2) + 0.3\cdot I(|i-j| =1)+I(|i-j|=0)$. Under the CS$(\rho)$ correlation structure, for the $i$th and $j$th factors, the correlation coefficient $corr = \rho^{I(i\neq j)}$. Under the categorical scenario, we first apply the same data generating approach as described above to obtain $\mathbf{U}$. Then for each $u_{i,j}$, the categorical measurement is generated as $I(u_{i,j}>-0.7)$. The threshold value $-0.7$ is chosen such that the proportion of $1$'s for each factor is roughly 75\%. Under each of the above simulation setting, consider the random error distribution $(1-\xi)N(0,1)+\xi Cauchy$, with the contamination probability $\xi=0$, 0.1, and 0.3. When $\xi=0$, the error distribution has no contamination and favors the nonrobust approaches, while the latter two values lead to different levels of contamination. The log event times are generated from the AFT model. The censoring times are generated independently from Weibull distributions. The censoring parameters are adjusted so that the censoring rates are about 25\%. Beyond the above scenarios, we also consider a set of parallel scenarios, under which there are 10 main E effects, 20 main G effects, and 40 interactions (that is, the number of important effects is doubled), and the nonzero coefficients are generated from $uniform(0.4, 0.6)$ (that is, the signal levels are reduced by about 50\%). Other settings remain the same.

The simulated data are analyzed using the proposed method. In addition, we also consider two alternatives: (a) the nonrobust method that adopts the weighted least squared loss and the same penalty as the proposed, and (b) the quantile regression-based method that adopts an $L_1$ robust loss and the same penalty as the proposed. We note that multiple other methods are potentially applicable. Comparing with the nonrobust method can directly establish the merit of being robust. The quantile regression-based approach is the most popular for high-dimensional data \citep{wu2014selective}. Thus these two alternatives are the most sensible to compare with.

All three methods involve tuning parameters. To eliminate the (possibly different) effects of tuning parameter selection on identification accuracy, we consider a sequence of tuning parameter values, evaluate identification accuracy at each value, and calculate the AUC (area under the ROC curve) as the overall measure. This approach has been adopted extensively in published studies \citep{zhu2014identifying}.

Summary statistics are computed based on 500 replicates. The AUC results for interactions and main effects combined are presented in Tables 1 and 2, respectively, for the scenarios with 35 and 70 important effects. To be thorough, we have also evaluated identification accuracy for interactions and main effects separately, and present the AUC results in Tables 4-7 in Appendix. For all three methods, the AUC value decreases as the contamination proportion increases, as expected. In Table 1, the proposed method outperforms the two alternatives under all except one scenario. In Table 2, it dominates the alternatives. Under some scenarios, the proposed method leads to a significant improvement in identification accuracy. For example in Table 1, with the continuous G distribution, 30\% contamination, and Band(0.3) correlation, the proposed method has a mean AUC of 0.901, while the alternatives have mean AUCs of 0.761 and 0.789. Compared to the nonrobust alternative, the proposed method also has smaller standard errors.

We have also experimented with a few other scenarios and made similar observations. In particular, we have examined the scenarios where the event and censoring times have weak to moderate correlations and observed similar satisfactory performance (details omitted). The proposed method and two alternatives respect the hierarchy. We have also looked into simpler alternatives, including MCP and Lasso, which may violate the hierarchy, and observed inferior performance.

\section{Analysis of the TCGA lung adenocarcinoma data}
\label{s:data}
Adenocarcinoma of the lung is the leading cause of cancer death worldwide. Profiling studies have been extensively conducted searching for its prognostic factors. Here we analyze the TCGA \citep{TCGA2014Comprehensivev} data on the prognosis of lung adenocarcinoma. The TCGA data were recently collected and published by NCI and have high quality. The prognosis outcome of interest is overall survival. The dataset contains measurements on 43 clinical/environmental variables and 18,897 gene expressions. There are a total of 468 patients, among whom 117 died during follow-up. The median follow-up time is 8 months. We select four E factors for downstream analysis, namely age, gender, smoking pack years, and smoking history. These factors have a relatively low missing rate in the TCGA dataset and have been previously suggested as potentially related to lung cancer prognosis. There are a total of 436 samples with both E and G measurements available. Among them, 110 died during follow-up, and the median follow-up time is 23 months. For the 326 censored subjects, the median follow-up time is 6 months. In principle, the proposed method can directly analyze all of the available gene expressions. To improve stability and reduce computational cost, we conduct marginal prescreening. Specifically, genes are screened based on their univariate regression significance (p-value less than or equal to 0.1) and interquartile range (above the median of all interquartile ranges). Similar prescreenings have been adopted in the literature. A total of 819 gene expressions are included in the downstream model fitting. Note that with the main G effects as well as interactions, the number of unknown parameters is much larger than the sample size.

Detailed estimation results are presented in Table 3 for the proposed method and Tables 8 and 9 in Appendix for the two alternatives. It is observed that the three methods lead to quite different findings. Specifically, the proposed and quantile methods share four common main G effects and four interactions. Otherwise, there is no overlap in identification. The ``signals'' in practical data can be weaker than those in simulated data, leading to the significant differences across methods.

With the proposed method, sixteen genes are identified to have interactions with either age or smoking status. As for many other cancer types, age has been identified as a critical factor in lung cancer prognosis. Smoking has been confirmed as the most important E factor for lung cancer risk and prognosis. In the literature, G-E interaction analysis for lung cancer prognosis is still very limited. However, there have been many studies on the functionalities of genes. Searching such studies can provide a partial support to the validity of our analysis results. Among the identified genes, many have been implicated in cancer in the literature. Specifically, the AGPAT family, which includes AGPAT6 as a member, has been found to play a role in multiple cancer types. For example, AGPAT2 and AGPAT11 have been found to be upregulated in ovarian, breast, cervical, and colorectal cancers \citep{agarwal2010enzymatic}. Another gene that is worth attention is ATF6, which acts both as a sensor and a transcription factor during endoplasmic reticulum stress. ATF6$\alpha$ has been found to promote hepatocarcinogenesis and cancer cell proliferation through activating downstream target gene BIP. Its efficiency of stress recognition and signaling has been found to decrease with age \citep{naidoo2009er}. We find that gene COLCA2 (colorectal cancer associated 2) interacts with smoking pack years. Studies have shown that COLCA2 may have critical functions in suppressing tumor formation in epithelial cells \citep{peltekova2014identification}. We also identify an interaction between NOS1AP and age. It has been found that the protein complex of SCRIB, NOS1AP, and VANGL1 regulates cell polarity and migration, and this complex can be associated with cancer progression \citep{anastas2012protein}. An interaction between PPP1R15B and smoking pack years has also been identified. It has been suggested that PPP1R15B is likely to be regulated by Nrf2, which has a protective response to smoking induced oxidative stress in the lung \citep{taylor2008network}. Also, PPP1R15B may promote cancer cell proliferation.

To complement the identification and estimation analysis, we also evaluate stability. Specifically, we randomly select 3/4 of the subjects and apply the proposed method and alternatives. This procedure is repeated 200 times. We then compute the probability that an interaction is identified. Similar procedures have been extensively adopted in published studies. The stability results are also provided in Tables 3, 8, and 9. We see that most of the identified interactions are relatively stable, with many having the probabilities of being identified close to one.

\section{Discussion}
To understand the prognosis of complex diseases, it is essential to study G-E interactions. In ``classic" low-dimensional biomedical studies, data contamination is found to be not rare, and it has been suggested that robust methods are needed to accommodate contamination. This study has developed a robust method for high-dimensional genetic interaction analysis, which is still limited in the literature.
The proposed method consists of a novel robust loss function and a penalized identification strategy that respects the ``main effects, interactions'' hierarchy, both of which have novel advancements. Also significantly advancing from the literature, we have rigorously established the consistency properties. The theoretical results may seem ``familiar'', which is ``comforting'' in that the consistency properties are not sacrificed with the additional robustness, high-dimensionality, and interactions. It is worth noting that the consistency results do not demand excessive assumptions on the error distribution, which are usually needed in the existing literature. In simulation, the proposed method outperforms the nonrobust alternative. It is interesting to note that it has superior performance when there is no contamination. Another important finding is that it also outperforms the quantile-based robust method. Most of the existing high-dimensional robust studies have adopted the quantile regression technique. Our simulation suggests that it is prudent to develop alternative robust methods. In the analysis of TCGA lung cancer data, the proposed method generates results with some overlappings with the quantile regression method, however, none with the nonrobust method. The identified genes have important implications, and the identified interactions are stable.

The proposed study can be potentially extended in multiple directions. In survival analysis, there are many other models beyond the AFT. It can be of interest to develop robust methods based on other models. We have studied G-E interactions. It can be of interest to extend to G-G interactions. In theoretical analysis, one problem left is the breakdown point. Because of the extremely high complexity, this problem has been left uninvestigated in many other robust studies too. In our simulation, we have experimented with contamination rate as high as 30\%, which is much higher than many of the existing studies. The superiority of the proposed method over the quantile regression method is observed. The relative efficiency of different robust methods, although of interest, will be postponed to future studies. In data analysis, the proposed method identifies a different set of main effects and interactions. Mining the literature and the stability evaluation can support the validity of findings to a certain extent. More validations need to be pursued in the future.

\section*{Acknowledgements}
This study was partly supported by the
MOE (Ministry of Education in China) Project of Humanities and Social Sciences (19YJC910010), Fundamental Research Funds for the Central Universities (20720171064 and 20720181003), and NIH (CA204120, CA196530).

\bibliographystyle{newapa}
\bibliography{reference}

\clearpage
\begin{table}
	\centering
	\caption{Simulation: identification of both G-E interactions and main G effects. In each cell, mean AUC (se). There are a total of 35 nonzero effects, with coefficients $\sim uniform(0.7, 1.3)$. }
	\label{tab1}
	\begin{tabular}{ccccc}
		\toprule
		$\xi$ & cor & proposed & nonrobust & quantile \\
		\midrule
		\multicolumn{5}{c}{continuous}\\
		\hline
		\multirow{6}{*}{0} & AR(0) & \textbf{0.891(0.065)} & 0.842(0.095) & 0.806(0.043) \\
		& AR(0.3) & \textbf{0.971(0.050)} & 0.917(0.096) & 0.832(0.025) \\
		& AR(0.8) & \textbf{0.981(0.041)} & 0.923(0.066) & 0.881(0.024) \\
		& Band(0.3) & \textbf{0.972(0.057)} & 0.908(0.106) & 0.828(0.024) \\
		& Band(0.6) & \textbf{0.978(0.044)} & 0.930(0.078) & 0.725(0.024) \\
		& CS(0.2) & \textbf{0.920(0.069)} & 0.827(0.096) & 0.854(0.024) \\
		\hline
		\multirow{6}{*}{0.1} & AR(0) & \textbf{0.824(0.077)} & 0.733(0.114) & 0.782(0.042) \\
		& AR(0.3) & \textbf{0.951(0.057)} & 0.858(0.130) & 0.815(0.031) \\
		& AR(0.8) & \textbf{0.970(0.061)} & 0.841(0.119) & 0.873(0.034) \\
		& Band(0.3) & \textbf{0.945(0.093)} & 0.850(0.143) & 0.802(0.021) \\
		& Band(0.6) & \textbf{0.959(0.058)} & 0.865(0.131) & 0.704(0.037) \\
		& CS(0.2) & \textbf{0.898(0.087)} & 0.779(0.109) & 0.846(0.043) \\
		\hline
		\multirow{6}{*}{0.3} & AR(0) & 0.769(0.086) & 0.646(0.097) & \textbf{0.775(0.045)} \\
		& AR(0.3) & \textbf{0.889(0.101)} & 0.742(0.147) & 0.788(0.021) \\
		& AR(0.8) &  \textbf{0.942(0.077)} & 0.754(0.124) & 0.856(0.025) \\
		& Band(0.3) &  \textbf{0.901(0.093)} & 0.761(0.140) & 0.789(0.022) \\
		& Band(0.6) &  \textbf{0.924(0.075)} & 0.785(0.131) & 0.691(0.041) \\
		& CS(0.2) &   \textbf{0.845(0.092)} & 0.661(0.117) & 0.831(0.045) \\
		\midrule
		\multicolumn{5}{c}{categorical}\\
		\hline
		\multirow{6}{*}{0} & AR(0) &   \textbf{0.890(0.062)} & 0.838(0.092) & 0.778(0.045) \\
		& AR(0.3) &   \textbf{0.963(0.054)} & 0.913(0.093) & 0.802(0.028) \\
		& AR(0.8) &   \textbf{0.975(0.041)} & 0.918(0.068) & 0.843(0.021) \\
		& Band(0.3) &   \textbf{0.971(0.041)} & 0.932(0.080) & 0.787(0.033) \\
		& Band(0.6) &   \textbf{0.972(0.039)} & 0.925(0.079) & 0.702(0.042) \\
		& CS(0.2) &   \textbf{0.917(0.082)} & 0.818(0.097) & 0.822(0.047) \\
		\hline
		\multirow{6}{*}{0.1} & AR(0) &   \textbf{0.835(0.085)} & 0.756(0.115) & 0.749(0.043) \\
		& AR(0.3) &   \textbf{0.944(0.055)} & 0.856(0.130) & 0.785(0.033) \\
		& AR(0.8) &   \textbf{0.970(0.037)} & 0.867(0.102) & 0.831(0.041) \\
		& Band(0.3) &   \textbf{0.953(0.052)} & 0.862(0.119) & 0.764(0.025) \\
		& Band(0.6) &   \textbf{0.965(0.044}) & 0.861(0.128) & 0.678(0.032) \\
		& CS(0.2) &   \textbf{0.895(0.086)} & 0.752(0.115) & 0.803(0.035) \\
		\hline
		\multirow{6}{*}{0.3} & AR(0) &   \textbf{0.771(0.090)} & 0.635(0.118) & 0.738(0.043) \\
		& AR(0.3) &   \textbf{0.895(0.087)} & 0.722(0.131) & 0.771(0.024) \\
		& AR(0.8) &   \textbf{0.946(0.057)} & 0.785(0.119) & 0.817(0.028) \\
		& Band(0.3) &   \textbf{0.897(0.115)} & 0.748(0.153) & 0.741(0.027) \\
		& Band(0.6) &   \textbf{0.921(0.083)} & 0.751(0.140) & 0.649(0.047) \\
		& CS(0.2) &   \textbf{0.822(0.110)} & 0.660(0.113) & 0.787(0.031) \\
		\bottomrule
	\end{tabular}
\end{table}

\clearpage

\begin{table}
	\centering
	\caption{Simulation: identification of both G-E interactions and main G effects. There are a total of 70 nonzero effects, with coefficients $\sim uniform(0.4, 0.6)$. }
	\label{tab2}
	\begin{tabular}{ccccc}
		\toprule
		$\xi$ & cor & proposed & nonrobust & quantile \\
		\midrule
		\multicolumn{5}{c}{continuous}\\
		\hline
		\multirow{6}{*}{0} & AR(0) & \textbf{0.678(0.043)} & 0.649(0.042) & 0.645(0.051) \\
		& AR(0.3) &  \textbf{0.800(0.045)} & 0.768(0.057) & 0.771(0.046) \\
		& AR(0.8) & \textbf{0.916(0.058)} & 0.828(0.054) & 0.812(0.044) \\
		& Band(0.3) & \textbf{0.827(0.057)} & 0.781(0.067)  & 0.787(0.039) \\
		& Band(0.6) & \textbf{0.865(0.052)} & 0.816(0.061) & 0.719(0.025) \\
		& CS(0.2) & \textbf{0.717(0.065)} & 0.645(0.047) & 0.668(0.037) \\
		\hline
		\multirow{6}{*}{0.1} & AR(0) & \textbf{0.651(0.040)} & 0.623(0.047) & 0.619(0.045) \\
		& AR(0.3) &  \textbf{0.737(0.069)} & 0.668(0.085) & 0.672(0.038) \\
		& AR(0.8) & \textbf{0.892(0.050)} & 0.779(0.081) & 0.795(0.047) \\
		& Band(0.3) & \textbf{0.790(0.061)} & 0.710(0.100) & 0.759(0.055) \\
		& Band(0.6) & \textbf{0.827(0.060)} & 0.767(0.080) & 0.754(0.041) \\
		& CS(0.2) & \textbf{0.691(0.061)} & 0.613(0.053) & 0.672(0.042) \\
		\hline
		\multirow{6}{*}{0.3} & AR(0) &  \textbf{0.605(0.052)} & 0.551(0.042) & 0.561(0.048) \\
		& AR(0.3) & \textbf{0.697(0.064)} & 0.601(0.058) & 0.633(0.037) \\
		& AR(0.8) &  \textbf{0.838(0.081)} & 0.679(0.093) & 0.719(0.042) \\
		& Band(0.3) &  \textbf{0.713(0.079)} & 0.608(0.080) & 0.668(0.039) \\
		& Band(0.6) &  \textbf{0.754(0.085)} & 0.648(0.102) & 0.651(0.035) \\
		& CS(0.2) &   \textbf{0.668(0.059)} & 0.568(0.065) & 0.611(0.041) \\
		\midrule
		\multicolumn{5}{c}{categorical}\\
		\hline
		\multirow{6}{*}{0} & AR(0) &  \textbf{0.675(0.045)} & 0.645(0.040) & 0.643(0.038) \\
		& AR(0.3) &   \textbf{0.784(0.057)} & 0.769(0.067) & 0.758(0.042) \\
		& AR(0.8) &   \textbf{0.909(0.058)} & 0.826(0.052) & 0.799(0.051) \\
		& Band(0.3) &   \textbf{0.799(0.058)} & 0.776(0.065) & 0.774(0.034) \\
		& Band(0.6) &   \textbf{0.847(0.063)} & 0.827(0.062) & 0.688(0.039) \\
		& CS(0.2) &   \textbf{0.719(0.064)} & 0.634(0.041) & 0.677(0.049) \\
		\hline
		\multirow{6}{*}{0.1} & AR(0) &   \textbf{0.654(0.052)} & 0.596(0.060)  & 0.604(0.037) \\
		& AR(0.3) &   \textbf{0.748(0.063)} & 0.683(0.093) & 0.695(0.052) \\
		& AR(0.8) &   \textbf{0.869(0.085)} & 0.764(0.086) & 0.772(0.041) \\
		& Band(0.3) &  \textbf{0.772(0.071)} & 0.712(0.093) & 0.733(0.039) \\
		& Band(0.6) &   \textbf{0.806(0.067)} & 0.736(0.099)  & 0.729(0.048) \\
		& CS(0.2) &  \textbf{0.684(0.058)} & 0.595(0.051) & 0.638(0.034) \\
		\hline
		\multirow{6}{*}{0.3} & AR(0) &   \textbf{0.614(0.056)} & 0.557(0.049) & 0.571(0.052) \\
		& AR(0.3) &   \textbf{0.697(0.065)} & 0.614(0.068) & 0.632(0.047) \\
		& AR(0.8) &  \textbf{0.824(0.092)} & 0.694(0.103)  & 0.727(0.045) \\
		& Band(0.3) &   \textbf{0.720(0.071)} & 0.639(0.085) & 0.655(0.036) \\
		& Band(0.6) &  \textbf{0.749(0.087)} & 0.644(0.090)  & 0.648(0.045) \\
		& CS(0.2) &   \textbf{0.666(0.056)} & 0.574(0.050) & 0.629(0.038) \\
		\bottomrule
	\end{tabular}
\end{table}

\clearpage
\begin{table}
	\centering
	\caption{Analysis of the TCGA lung adenocarcinoma data using the proposed method. The identified interactions are denoted as ``gene * environmental variable". For the interactions, values in ``()" are the stability results.}
	\label{tab4}
	\begin{tabular}{lr}
		\toprule
		effect & estimate $\times$100\\
		\midrule
		age & 8.817  \\
		smoking pack years & -0.358  \\
		AGPAT6 & 55.343  \\
		ANKRD46 & 4.646   \\
		ATF6 & 40.350   \\
		C1ORF27 & 7.708   \\
		COLCA2 & -1.808   \\
		CAND1 & 32.138   \\
		DNAJC21 & 6.652   \\
		DYRK2 & -24.595   \\
		HERPUD2 & -40.358   \\
		LCMT2 & 40.151   \\
		NOS1AP & -28.707   \\
		PIGZ & -19.058   \\
		PPP1R15B & -2.411   \\
		TROVE2 & -5.979   \\
		WIPI2 & -18.739   \\
		YTHDF3 & 21.524   \\
		AGPAT6 * age & 0.202(0.995) \\
		ANKRD46 * age & 0.546(0.537) \\
		ATF6 * age & 0.493(0.193) \\
		C1ORF27 * age & -0.072(0.989) \\
		COLCA2 * smoking pack years & 0.315(0.993) \\
		CAND1 * smoking pack years & -0.716(0.649) \\
		DNAJC21 * age & -0.280(1.000) \\
		DYRK2 * age & 0.496(0.330) \\
		HERPUD2 * age & -0.393(0.975) \\
		LCMT2 * age & -0.222(0.927) \\
		NOS1AP * age & 0.140(0.734) \\
		PIGZ * age & 0.046(0.397) \\
		PPP1R15B * smoking pack years & 0.711(0.998) \\
		TROVE2 * age & -0.416(0.890) \\
		WIPI2 * age & -0.205(1.000) \\
		YTHDF3 * age & -0.201(0.839) \\
		\bottomrule
	\end{tabular}
\end{table}

\clearpage

\section*{Appendix}

\textbf{Proof for Theorem 1}

\begin{proof} Define the oracle estimator  $\widehat{\bm\zeta}$ with $\widehat{\bm\zeta}_{\mathcal{A}^c}=0$ and
	\begin{equation}\label{oracle}
	\widehat{\bm\zeta}_\mathcal{A} = {\arg\max} \sum_{i=1}^n\omega_i\exp(-(y_i - \mathbf{u}_{i, \mathcal{A}}^\top{\bm\zeta}_\mathcal{A})^2 / \theta).
	\end{equation}
	Recall  that the proposed objective function is
	\begin{equation}\label{app1}
L_{\lambda_1, \lambda_2, \theta}(\bm\zeta)=Q_{\theta}(\bm\zeta)-\sum_{k=1}^{p}\rho(\|b_k\|; \lambda_1, s)-\sum_{k=1}^{p}\sum_{j=2}^{q+1}\rho(|b_{kj}|; \lambda_2, s).	\end{equation}
In what follows, we first establish the estimation consistency of $\widehat{\bm\zeta}$ in Step 1, and then show that $\widehat{\bm\zeta}$ is a local maximizer of $L_{\lambda_1, \lambda_2, \theta}(\bm\zeta)$ is Step 2.

\underline{\textbf{{ Step 1.}}}
Define the objective function
\begin{equation*}
R_n({\bm\zeta}_\mathcal{A}) = \sum_{i=1}^n\omega_i\exp(-(y_i - \mathbf{u}_{i, \mathcal{A}}^\top{\bm\zeta}_\mathcal{A})^2 / \theta).
\end{equation*}
Then $\widehat{\bm\zeta}_\mathcal{A} = {\arg\max}R_n({\bm\zeta}_\mathcal{A})$. Let $r_n = \sqrt{|\mathcal{A}|/n}$. To prove $\|\widehat{\bm\zeta}_\mathcal{A} - {\bm\zeta}_\mathcal{A}^*\|_2 = O_p(r_n)$, it suffices to show that for any given $\eta>0$, there exists a sufficiently large constant $C>0$,
	\begin{equation}\label{tail0}
		\Pr\Big(\sup\limits_{
			{\bm\zeta}_\mathcal{A}\in \mathcal{I}}R_n({\bm\zeta}_\mathcal{A})<R_n({\bm\zeta}_\mathcal{A}^*) \Big)\geq 1-\eta,
	\end{equation}
	where $\mathcal{I}=\left\{{\bm\zeta}_\mathcal{A}: \|{\bm\zeta}_\mathcal{A} - {\bm\zeta}_\mathcal{A}^*\|_2 = Cr_n\right\}$.
This implies that $R_n({\bm\zeta}_\mathcal{A})$ has a local maximizer $\widehat {\bm\zeta}_\mathcal{A}$ that satisfies
$\|{\bm\zeta}_\mathcal{A} - {\bm\zeta}_\mathcal{A}^*\|_2 = O_p(r_n)$.

Recall the definitions of $D_{n}({\bm\zeta})$ and $I_{n}({\bm\zeta})$. By Taylor's expansion, we have
	\begin{eqnarray}\label{q0}
		R_n({\bm\zeta}_\mathcal{A})-R_n({\bm\zeta}_\mathcal{A}^*)
		&=& \sum_{i=1}^n\omega_i\left\{\exp(-(y_i - \mathbf{u}_{i, \mathcal{A}}^\top{\bm\zeta}_\mathcal{A})^2 / \theta)-\exp(-(y_i - \mathbf{u}_{i, \mathcal{A}}^\top{\bm\zeta}_\mathcal{A}^*)^2 / \theta)\right\} \nonumber\\
		&=& D_{n, \mathcal{A}}({\bm\zeta}^*)^\top ({\bm\zeta}_\mathcal{A} - {\bm\zeta}_\mathcal{A}^*)+ \frac{1}{2}({\bm\zeta}_\mathcal{A} - {\bm\zeta}_\mathcal{A}^*)^\top I_{n,{\mathcal{A}\mathcal{A}}}(\bar{\bm\zeta})({\bm\zeta}_\mathcal{A} - {\bm\zeta}_\mathcal{A}^*) \nonumber\\
		&\dot{= }& Q_1 + Q_2 ,
	\end{eqnarray}
	where $\bar{\bm\zeta}$ lies between ${\bm\zeta}^*$ and ${\bm\zeta}$.
 By C3 and C4, we have that for all $j\in \{1, \cdots, p+q+pq+1\}$ and any given $t$, $\Pr(|D_{n,j}({\bm\zeta}^*)| >t )\leq 2 \exp\left(- n t^2/\sigma^2\right)$.
  Then $E(|\sqrt{n}D_{n,j}({\bm\zeta}^*)|)<K<\infty$ for all $j$. With the Markov's inequality,
\begin{eqnarray*}
\Pr(\|D_{n, \mathcal{A}}({\bm\zeta}^*)\|_2>t ) \leq E[\|\sqrt{n}D_{n, \mathcal{A}}({\bm\zeta}^*)\|_2^2]/(n t^2)  \leq  |\mathcal{A}|K/(n t^2) .
\end{eqnarray*}
 By the Cauchy-Schwarz inequality, $Q_1\leq C\|D_{n, \mathcal{A}}({\bm\zeta}^*)\|_2 r_n$. Let $t = C\rho_* r_n/3$, where $\rho_*$ is the smallest eigenvalue of $-I_{AA}({\bm\zeta}_A^*)$. From C5, we have that $\rho_*$ is bounded away from zero and infinity.  Then we have
\begin{eqnarray}\label{q1}
\Pr(Q_1 \leq \frac{1}{3}\rho_* C^{2}r_n^2)\leq  1- \frac{9K}{C^2\rho^2_*}.
\end{eqnarray}

 For $Q_2$, we have
	\begin{eqnarray}\label{q2-0}
2Q_2&=&({\bm\zeta}_\mathcal{A} - {\bm\zeta}_\mathcal{A}^*)^\top I_{{\mathcal{A}\mathcal{A}}}({\bm\zeta}^*)({\bm\zeta}_\mathcal{A} - {\bm\zeta}_\mathcal{A}^*) +  ({\bm\zeta}_\mathcal{A} - {\bm\zeta}_\mathcal{A}^*)^\top \left\{I_{{\mathcal{A}\mathcal{A}}}(\bar{{\bm\zeta}})-I_{{\mathcal{A}\mathcal{A}}}({\bm\zeta}^*)\right\}
({\bm\zeta}_\mathcal{A} - {\bm\zeta}_\mathcal{A}^*)\nonumber\\
&& + ({\bm\zeta}_\mathcal{A} - {\bm\zeta}_\mathcal{A}^*)^\top \left\{I_{n,{\mathcal{A}\mathcal{A}}}(\bar{{\bm\zeta}})-I_{{\mathcal{A}\mathcal{A}}}(\bar{{\bm\zeta}})\right\}
		({\bm\zeta}_\mathcal{A} - {\bm\zeta}_\mathcal{A}^*)\nonumber\\
		&\dot{= }& Q_{21} + Q_{22} +Q_{23}.
	\end{eqnarray}
Since $\lambda_{\max}(I_{{\mathcal{A}\mathcal{A}}}(\bm\zeta^*))\leq -\rho_*$ by C5, we have 	
\begin{eqnarray}\label{q2-1}
Q_{21} \leq - \rho_* C^2 r_n^2.
\end{eqnarray}
Under C4, we have
\begin{eqnarray}\label{q2-2}
Q_{22} \leq \kappa \|\bar{{\bm\zeta}}-{\bm\zeta}^*\|_2  C^2 r_n^2 < \kappa  C^3 r_n^3< \frac{1}{6}  C^2 \rho_* r_n^2.
\end{eqnarray}
The secend inequality holds since $\bar{\bm\zeta}$ lies between ${\bm\zeta}^*$ and ${\bm\zeta}$, which yields $\|\bar{{\bm\zeta}}-{\bm\zeta}^*\|_2<C r_n$. When $n$ is sufficiently large, the last inequality holds.
With C4 and the Bonferroni's inequality,
	 \[\Pr(\|I_{n,{\mathcal{A}\mathcal{A}}}(\bar{\bm\zeta})-
I_{{\mathcal{A}\mathcal{A}}}(\bar{\bm\zeta})\|_F^2 \geq \rho_*^2/9 ) \leq 2|\mathcal{A}|^2 \exp\left(-  n \rho_*^2/\sigma^2\right),\]
where $\|\cdot\|_F$ denotes the Frobenius norm. By the inequality $\lambda_{\max}(I_{n,{\mathcal{A}\mathcal{A}}}(\bar{{\bm\zeta}})
-I_{{\mathcal{A}\mathcal{A}}}(\bar{{\bm\zeta}}))\leq \|I_{n,{\mathcal{A}\mathcal{A}}}(\bm\zeta^*)-I_{{\mathcal{A}\mathcal{A}}}
({\bm\zeta}^*)\|_F$, we have
\begin{eqnarray}\label{q2-3}
Q_{23}\leq \frac{1}{3} \rho_*  C^2  r_n^2 ~ \mbox{with probability at least} ~~1-2|\mathcal{A}|^2 \exp\left(-  n \rho_*^2/\sigma^2\right).
\end{eqnarray}
 Combining (\ref{q2-0}), (\ref{q2-1}), (\ref{q2-2}), and (\ref{q2-3}), we have
\begin{eqnarray}\label{q2}
\Pr(Q_{2} <-\frac{1}{2} \rho_* C^2  r_n^2)\geq 1-2|\mathcal{A}|^2 \exp\left(- n \rho_*^2/\sigma^2\right).
\end{eqnarray}

With (\ref{q0}), (\ref{q1}), and (\ref{q2}), we have
\begin{eqnarray}
	R_n({\bm\zeta}_\mathcal{A})-R_n({\bm\zeta}_\mathcal{A}^*)< -\frac{1}{6} \rho_*  C^2 r_n^2 <0
\end{eqnarray}
with probability at least
\[1- \frac{9K}{C^2\rho^2_*}-  2|\mathcal{A}|^2 \exp\left(- n \rho_*^2/\sigma^2\right).\]
Note that $\rho_*$ is bounded away from zero and infinity in C5. As $n\rightarrow \infty$, the above probability is bigger than $1- \frac{16K}{C^2\rho^2_*}$. Let $C=4\rho_*^{-1} \sqrt{K/\eta}$, then we can conclude \eqref{tail0}.

\underline{\textbf{{ Step 2.}}} Next we show that the oracle estimator $\widehat{\bm\zeta}$ studied in Step 1  satisfies the Karush-Kuhn-Tucher (KKT) condition, and then  $\widehat{\bm\zeta}$ is a local maximizer of $L_{\lambda_1, \lambda_2, \theta}(\bm\zeta)$. Based on the results in Step 1 and C6, we only need to check the following conditions
\begin{equation}\label{kkt}
\left\|D_{n, \mathcal{B}_1}(\widehat{\bm\zeta})\right\|_\infty<\lambda_1,~ \left\|D_{n, \mathcal{B}_2}(\widehat{\bm\zeta})\right\|_\infty<\lambda_2, ~\left\|D_{n, \mathcal{B}_3}(\widehat{\bm\zeta})\right\|_\infty<\lambda_1 + \lambda_2
\end{equation}
hold with asymptotic probability one,  where $\|\nu\|_\infty = \max_i |\nu_i|$ for any vector $\nu= (\nu_1, \cdots, \nu_{|\mathcal{A}^c|})$.
Applying the Taylor's expansion,
	\begin{eqnarray}\label{omega}
		D_{n, \mathcal{B}_1}(\widehat{\bm\zeta})&=& D_{n, \mathcal{B}_1}({\bm\zeta}^*)+ I_{n,\mathcal{B}_1\mathcal{A}}(\widetilde{\bm\zeta})(\widehat{\bm\zeta}_\mathcal{A} - {\bm\zeta}_\mathcal{A}^* ),
	\end{eqnarray}
	where $\widetilde{\bm\zeta}$ lies between ${\bm\zeta}^*$ and $\widehat{{\bm\zeta}}$. From (\ref{oracle}) and the proof of Theorem 1(a), we have
	\begin{eqnarray}\label{hatzeta}
		\widehat{\bm\zeta}_\mathcal{A} - {\bm\zeta}_\mathcal{A}^* = -I_{n, {\mathcal{A}\mathcal{A}}}(\bar{{\bm\zeta}})^{-1} D_{n, \mathcal{A}}({\bm\zeta}^*),
	\end{eqnarray}
	where $\bar{\bm\zeta}$ lies between ${\bm\zeta}^*$ and $\widehat{{\bm\zeta}}$, which is defined in Step 1. By substituting (\ref{hatzeta}) into (\ref{omega}),
	\begin{eqnarray}\label{delta}
		D_{n, \mathcal{B}_1}(\widehat{\bm\zeta})= D_{n, \mathcal{B}_1}({\bm\zeta}^*) - I_{n,\mathcal{B}_1 \mathcal{A}}(\widetilde{\bm\zeta})I_{n, {\mathcal{A}\mathcal{A}}}(\bar{{\bm\zeta}})^{-1} D_{n, \mathcal{A}}({\bm\zeta}^*).
	\end{eqnarray}
Here we define
	\begin{eqnarray*}
		\Delta_{n, \mathcal{B}_1}^*
		=  D_{n, \mathcal{B}_1}({\bm\zeta}^*) - I_{\mathcal{B}_1 \mathcal{A}}({\bm\zeta}^*)I_{ {\mathcal{A}\mathcal{A}}}({{\bm\zeta}}^*)^{-1} D_{n, \mathcal{A}}({\bm\zeta}^*).
	\end{eqnarray*}
Inspired by the deduction of $Q_2$ in Step 1, we can establish that \[\Pr(\|D_{n, \mathcal{B}_1}(\widehat{\bm\zeta})\|_\infty>\lambda_1)\asymp \Pr(\|\Delta_{n, \mathcal{B}_1}^*\|_\infty>\lambda_1).\] That is, we only need to focus on $\|\Delta_{n, \mathcal{B}_1}^*\|_\infty$ in order to evaluate the probability of $\{\|D_{n, \mathcal{B}_1}(\widehat{\bm\zeta})\|_\infty<\lambda_1\}$ in (\ref{kkt}). Note that,
	\begin{eqnarray}\label{ome}
\|\Delta_{n, \mathcal{B}_1}^*\|_\infty
		&\leq & \|D_{n,\mathcal{B}_1}({\bm\zeta}^*)\|_\infty+\|I_{\mathcal{B}_1 \mathcal{A}}({\bm\zeta}^*)I_{ {\mathcal{A}\mathcal{A}}}({{\bm\zeta}}^*)^{-1} D_{n, \mathcal{A}}({\bm\zeta}^*)\|_\infty\nonumber\\
		&\leq & \|D_{n}({\bm\zeta}^*)\|_\infty + \|I_{\mathcal{B}_1 \mathcal{A}}({\bm\zeta}^*)I_{ {\mathcal{A}\mathcal{A}}}({{\bm\zeta}}^*)^{-1}\|_\infty \|D_{n}({\bm\zeta}^*)\|_\infty.
	\end{eqnarray}
	Recall that $\Phi_1= \|I_{\mathcal{B}_1 \mathcal{A}}({\bm\zeta}^*)I_{ {\mathcal{A}\mathcal{A}}}({{\bm\zeta}}^*)^{-1}\|_\infty$. If
	\begin{eqnarray*}
		\|D_{n}({\bm\zeta}^*)\|_\infty <\frac{\lambda_1 }{1+\Phi_1},
	\end{eqnarray*} along with (\ref{ome}), we have $\|\Delta_{n, \mathcal{B}_1}^*\|_\infty <\lambda_1$. Similarly, we also need
\begin{eqnarray*}
\|D_{n}({\bm\zeta}^*)\|_\infty <\frac{\lambda_2}{1+\Phi_2},~\mbox{and}~ \|D_{n}({\bm\zeta}^*)\|_\infty <\frac{\lambda_1+\lambda_2}{1+\Phi_3}
\end{eqnarray*}
to satisfy the other two conditions in  (\ref{kkt}), where $\Phi_2= \|I_{\mathcal{B}_2 \mathcal{A}}({\bm\zeta}^*)I_{ {\mathcal{A}\mathcal{A}}}({{\bm\zeta}}^*)^{-1}\|_\infty$ and $\Phi_3= \|I_{\mathcal{B}_3 \mathcal{A}}({\bm\zeta}^*)I_{ {\mathcal{A}\mathcal{A}}}({{\bm\zeta}}^*)^{-1}\|_\infty$. Based on the above discussions, we have
\[\|D_{n}({\bm\zeta}^*)\|_\infty < \frac{\lambda_1\wedge \lambda_2}{1+\max_{t=1}^3{\Phi_t}}< \min\{\frac{\lambda_1 }{1+\Phi_1}, \frac{\lambda_2}{1+\Phi_2},\frac{\lambda_1+\lambda_2}{1+\Phi_3} \}.\]
 We now derive the probability bound for the above event. By the Bonferroni's inequality and C4, we can obtain
	\begin{eqnarray*}
		\Pr\left\{\|D_{n}({\bm\zeta}^*)\|_\infty <  \frac{\lambda_1\wedge \lambda_2}{1+\max_{t=1}^3{\Phi_t}}\right\}\geq 1- 2(pq+p+q+1)\exp\left(-\frac{ n (\lambda_1\wedge \lambda_2)^2}{(1+\max_{t=1}^3{\Phi_t})^2\sigma^2}\right).
	\end{eqnarray*}

Combining the results in Steps 1 and 2, we conclude that  $\widehat{\bm\zeta}$ is a local maximizer of  $L_{\lambda_1, \lambda_2, \theta}(\bm\zeta)$ with probability at least $$ 1-O\left(p\exp\left(-\frac{n (\lambda_1\wedge \lambda_2)^2}{(1+\max_{t=1}^3{\Phi_t})^2\sigma^2 }\right)\right),$$
	and satisfies $\|\widehat{\bm\zeta}_{\mathcal{A}} - {\bm\zeta}^*_{\mathcal{A}}\|_2=O_p(\sqrt{|\mathcal{A}|/n}),~ \widehat{\bm\zeta}_{\mathcal{A}^c}=0$.
With C6,  $\log p = O ( n \varpi_n^2)$, and $\varpi_n= (\lambda_1\wedge \lambda_2)/\{\max(\Phi_1, \Phi_2, \Phi_3)\}$, this tail probability is exponentially small.
The theorem is thus proved.

\end{proof}

\clearpage

\begin{table}
	\centering
	\caption{
Simulation: identification of main G effects. In each cell, mean AUC (se). There are a total of 10 nonzero main effects, with coefficients $\sim uniform(0.7, 1.3)$. }
	\label{tab3}
	\begin{tabular}{ccccc}
		\toprule
		$\xi$ & cor & robust & nonrobust & quantile \\
		\midrule
		\multicolumn{5}{c}{continuous}\\
		\hline
		\multirow{6}{*}{0} & AR(0) & 0.94(0.055) & \textbf{0.964(0.052)} & 0.867(0.041) \\
   & AR(0.3) & 0.985(0.019) &  \textbf{0.998(0.002)} & 0.882(0.052) \\
   & AR(0.8) & 0.987(0.019) &  \textbf{0.994(0.019)} & 0.912(0.032) \\
   & BAND(0.3) & 0.985(0.020) &  \textbf{0.999(0.003)} & 0.841(0.046) \\
   & BAND(0.6) & 0.985(0.022) &  \textbf{0.998(0.007)} & 0.792(0.047) \\
   & CS(0.2) &  \textbf{0.938(0.047)} & 0.923(0.061) & 0.863(0.039) \\
		\hline
		\multirow{6}{*}{0.1} & AR(0) &  \textbf{0.883(0.079)} & 0.834(0.135) & 0.852(0.044) \\
   & AR(0.3) &  \textbf{0.975(0.028)} & 0.956(0.108) & 0.841(0.053) \\
   & AR(0.8) &  \textbf{0.981(0.052)} & 0.922(0.127) & 0.891(0.034) \\
   & BAND(0.3) &  \textbf{0.967(0.075)} & 0.942(0.136) & 0.836(0.036) \\
   & BAND(0.6) &  \textbf{0.975(0.036)} & 0.944(0.128) & 0.789(0.044) \\
   & CS(0.2) &  \textbf{0.921(0.079)} & 0.851(0.126) & 0.855(0.031) \\
		\hline
		\multirow{6}{*}{0.3} & AR(0) &  \textbf{0.837(0.102)} & 0.716(0.127) & 0.792(0.028) \\
   & AR(0.3) &  \textbf{0.942(0.082)} & 0.829(0.168) & 0.814(0.031) \\
   & AR(0.8) &  \textbf{0.970(0.070)} & 0.841(0.135) & 0.855(0.042) \\
   & BAND(0.3) &  \textbf{0.943(0.079)} & 0.836(0.158) & 0.821(0.028) \\
   & BAND(0.6) &  \textbf{0.954(0.067)} & 0.871(0.144) & 0.811(0.053) \\
   & CS(0.2) &  \textbf{0.886(0.095)} & 0.704(0.142) & 0.824(0.058) \\
		\midrule
		\multicolumn{5}{c}{categorical}\\
		\hline
		\multirow{6}{*}{0} & AR(0) & 0.931(0.047) &  \textbf{0.956(0.050)} & 0.857(0.044) \\
   & AR(0.3) & 0.970(0.030) &  \textbf{0.998(0.010)} & 0.872(0.058) \\
   & AR(0.8) & 0.976(0.028) &  \textbf{0.992(0.023)} & 0.883(0.044) \\
   & BAND(0.3) & 0.974(0.027) &  \textbf{0.999(0.002)} & 0.832(0.041) \\
   & BAND(0.6) & 0.974(0.029) &  \textbf{0.999(0.010)} & 0.824(0.036) \\
   & CS(0.2) &  \textbf{0.937(0.051)} & 0.918(0.065) & 0.844(0.051) \\
		\hline
		\multirow{6}{*}{0.1} & AR(0) &   \textbf{0.894(0.078)} & 0.868(0.137) & 0.853(0.039) \\
   & AR(0.3) &  \textbf{0.963(0.036)} & 0.959(0.104) & 0.875(0.056) \\
   & AR(0.8) &  \textbf{0.975(0.029)} & 0.939(0.090) & 0.897(0.064) \\
   & BAND(0.3) &  \textbf{0.968(0.036)} & 0.960(0.090) & 0.841(0.042) \\
   & BAND(0.6) &  \textbf{0.971(0.033)} & 0.934(0.117) & 0.829(0.048) \\
   & CS(0.2) &  \textbf{0.923(0.075)} & 0.824(0.138) & 0.846(0.057) \\
		\hline
		\multirow{6}{*}{0.3} & AR(0) &  \textbf{0.838(0.098)} & 0.707(0.153) & 0.788(0.029) \\
   & AR(0.3) &  \textbf{0.932(0.075)} & 0.808(0.159) & 0.818(0.035) \\
   & AR(0.8) &  \textbf{0.967(0.041)} & 0.863(0.143) & 0.854(0.048) \\
   & BAND(0.3) &  \textbf{0.931(0.101)} & 0.842(0.173) & 0.819(0.035) \\
   & BAND(0.6) &  \textbf{0.946(0.065)} & 0.821(0.155) & 0.813(0.044) \\
   & CS(0.2) &  \textbf{0.854(0.106)} & 0.704(0.136) & 0.828(0.052) \\
		\bottomrule
	\end{tabular}
\end{table}

\clearpage
\begin{table}
	\centering
	\caption{Simulation: identification of G-E interactions. In each cell, mean AUC (se). There are a total of 20 nonzero interactions, with coefficients $\sim uniform(0.7, 1.3)$. }
	\label{tab4}
	\begin{tabular}{ccccc}
		\toprule
		$\xi$ & cor & robust & nonrobust & quantile \\
		\midrule
		\multicolumn{5}{c}{continuous}\\
		\hline
		\multirow{6}{*}{0} & AR(0) &  \textbf{0.866(0.088)} & 0.784(0.137) & 0.761(0.047) \\
   & AR(0.3) &  \textbf{0.963(0.075)} & 0.873(0.143) & 0.776(0.062) \\
   & AR(0.8) &  \textbf{0.976(0.061)} & 0.888(0.098) & 0.862(0.066) \\
   & BAND(0.3) &  \textbf{0.964(0.085)} & 0.861(0.159) & 0.803(0.058) \\
   & BAND(0.6) &  \textbf{0.973(0.066)} & 0.895(0.117) & 0.659(0.043) \\
   & CS(0.2) &  \textbf{0.904(0.092)} & 0.775(0.134) & 0.848(0.039) \\
		\hline
		\multirow{6}{*}{0.1} & AR(0) &  \textbf{0.792(0.090)} & 0.684(0.124) & 0.734(0.053) \\
   & AR(0.3) &  \textbf{0.938(0.082)} & 0.811(0.158) & 0.786(0.048) \\
   & AR(0.8) &  \textbf{0.962(0.075)} & 0.801(0.128) & 0.864(0.039) \\
   & BAND(0.3) &  \textbf{0.932(0.112)} & 0.806(0.165) & 0.775(0.041) \\
   & BAND(0.6) &  \textbf{0.948(0.080)} & 0.826(0.150) & 0.633(0.052) \\
   & CS(0.2) &  \textbf{0.881(0.105)} & 0.740(0.130) & 0.829(0.058) \\
		\hline
		\multirow{6}{*}{0.3} & AR(0) & 0.733(0.094) & 0.612(0.095) &  \textbf{0.753(0.046)} \\
   & AR(0.3) &  \textbf{0.862(0.122)} & 0.701(0.152) & 0.749(0.052) \\
   & AR(0.8) &  \textbf{0.927(0.089)} & 0.710(0.128) & 0.861(0.057) \\
   & BAND(0.3) &  \textbf{0.879(0.110)} & 0.725(0.146) & 0.748(0.033) \\
   & BAND(0.6) &  \textbf{0.907(0.089)} & 0.744(0.139) & 0.598(0.062) \\
   & CS(0.2) & 0.820(0.105) & 0.637(0.114) &  \textbf{0.841(0.045)} \\
		\midrule
		\multicolumn{5}{c}{categorical}\\
		\hline
		\multirow{6}{*}{0} & AR(0) &  \textbf{0.866(0.086)} & 0.782(0.130) & 0.733(0.064) \\
   & AR(0.3) &  \textbf{0.955(0.079)} & 0.869(0.140) & 0.728(0.051) \\
   & AR(0.8) &  \textbf{0.971(0.061)} & 0.881(0.100) & 0.802(0.039) \\
   & BAND(0.3) &  \textbf{0.967(0.061)} & 0.898(0.119) & 0.749(0.048) \\
   & BAND(0.6) &  \textbf{0.969(0.058)} & 0.888(0.119) & 0.609(0.051) \\
   & CS(0.2) &  \textbf{0.900(0.109)} & 0.763(0.134) & 0.801(0.039) \\
		\hline
		\multirow{6}{*}{0.1} & AR(0) &  \textbf{0.801(0.102)} & 0.702(0.125) & 0.667(0.057) \\
   & AR(0.3) &  \textbf{0.932(0.077)} & 0.806(0.159) & 0.702(0.048) \\
   & AR(0.8) &  \textbf{0.964(0.056)} & 0.830(0.123) & 0.764(0.055) \\
   & BAND(0.3) &  \textbf{0.942(0.074)} & 0.814(0.151) & 0.689(0.053) \\
   & BAND(0.6) &  \textbf{0.959(0.064)} & 0.826(0.148) & 0.604(0.045) \\
   & CS(0.2) &  \textbf{0.875(0.104)} & 0.713(0.123) & 0.751(0.048) \\
		\hline
		\multirow{6}{*}{0.3} & AR(0) &  \textbf{0.734(0.099)} & 0.601(0.114) & 0.681(0.039) \\
   & AR(0.3) &  \textbf{0.873(0.104)} & 0.681(0.131) & 0.687(0.058) \\
   & AR(0.8) &  \textbf{0.931(0.075)} & 0.746(0.120) & 0.768(0.048) \\
   & BAND(0.3) &  \textbf{0.877(0.132)} & 0.704(0.157) & 0.699(0.059) \\
   & BAND(0.6) &  \textbf{0.905(0.101)} & 0.718(0.146) & 0.547(0.067) \\
   & CS(0.2) &  \textbf{0.800(0.124)} & 0.635(0.115) & 0.724(0.055) \\
		\bottomrule
	\end{tabular}
\end{table}

\clearpage
\begin{table}
	\centering
	\caption{Simulation: identification of main G effects. In each cell, mean AUC (se). There are a total of 20 nonzero main effects, with coefficients $\sim uniform(0.4, 0.6)$. }
	\label{tab5}
	\begin{tabular}{ccccc}
		\toprule
		$\xi$ & cor & robust & nonrobust & quantile \\
		\midrule
		\multicolumn{5}{c}{continuous}\\
		\hline
		\multirow{6}{*}{0} & AR(0) & 0.735(0.065) & \textbf{0.738(0.057)} & 0.684(0.042) \\
   & AR(0.3) & 0.885(0.049) & \textbf{0.894(0.046)} & 0.798(0.038) \\
   & AR(0.8) & \textbf{0.961(0.045)} & 0.925(0.036) & 0.811(0.048) \\
   & BAND(0.3) & \textbf{0.896(0.046)} & 0.894(0.047) & 0.809(0.044) \\
   & BAND(0.6) & 0.916(0.050) & \textbf{0.921(0.047)} & 0.792(0.039) \\
   & CS(0.2) & \textbf{0.769(0.065)} & 0.710(0.059) & 0.753(0.049) \\
		\hline
		\multirow{6}{*}{0.1} & AR(0) & \textbf{0.701(0.064)} & 0.682(0.071) & 0.678(0.033) \\
   & AR(0.3) & \textbf{0.806(0.079)} & 0.754(0.112) & 0.794(0.049) \\
   & AR(0.8) & \textbf{0.947(0.046)} & 0.871(0.088) & 0.806(0.053) \\
   & BAND(0.3) & \textbf{0.865(0.071)} & 0.809(0.132) & 0.801(0.036) \\
   & BAND(0.6) & \textbf{0.886(0.064)} & 0.860(0.086) & 0.789(0.055) \\
   & CS(0.2) & 0.738(0.080) & 0.659(0.084) & \textbf{0.784(0.042)} \\
		\hline
		\multirow{6}{*}{0.3} & AR(0) & \textbf{0.646(0.073)} & 0.577(0.067) & 0.632(0.044) \\
   & AR(0.3) & \textbf{0.774(0.090)} & 0.664(0.088) & 0.672(0.052) \\
   & AR(0.8) & \textbf{0.896(0.081)} & 0.743(0.127) & 0.755(0.041) \\
   & BAND(0.3) & \textbf{0.782(0.106)} & 0.664(0.112) & 0.711(0.065) \\
   & BAND(0.6) & \textbf{0.823(0.104)} & 0.719(0.141) & 0.705(0.053) \\
   & CS(0.2) & \textbf{0.708(0.078)} & 0.601(0.093) & 0.645(0.051) \\
   \midrule
		\multicolumn{5}{c}{categorical}\\
		\hline
		\multirow{6}{*}{0} & AR(0) & \textbf{0.736(0.068)} & 0.729(0.054) & 0.679(0.045) \\
   & AR(0.3) & 0.858(0.066) & \textbf{0.901(0.059)} & 0.782(0.052) \\
   & AR(0.8) & \textbf{0.944(0.055)} & 0.922(0.046) & 0.797(0.041) \\
   & BAND(0.3) & 0.869(0.062) & \textbf{0.900(0.057)} & 0.787(0.058) \\
   & BAND(0.6) & 0.894(0.062) & \textbf{0.920(0.045)} & 0.762(0.048) \\
   & CS(0.2) & \textbf{0.768(0.067)} & 0.696(0.055) & 0.763(0.048) \\
		\hline
		\multirow{6}{*}{0.1} & AR(0) & \textbf{0.716(0.074)} & 0.659(0.096) & 0.669(0.051) \\
   & AR(0.3) & \textbf{0.826(0.077)} & 0.777(0.132) & 0.752(0.039) \\
   & AR(0.8) & \textbf{0.914(0.089)} & 0.837(0.103) & 0.786(0.054) \\
   & BAND(0.3) & \textbf{0.838(0.079)} & 0.805(0.120) & 0.743(0.062) \\
   & BAND(0.6) & \textbf{0.867(0.069)} & 0.828(0.118) & 0.721(0.039) \\
   & CS(0.2) & \textbf{0.723(0.070)} & 0.642(0.080) & 0.678(0.042) \\
		\hline
		\multirow{6}{*}{0.3} & AR(0) & \textbf{0.639(0.074)} & 0.588(0.074) & 0.613(0.045) \\
   & AR(0.3) & \textbf{0.758(0.083)} & 0.684(0.097) & 0.658(0.047) \\
   & AR(0.8) & \textbf{0.877(0.109)} & 0.764(0.138) & 0.743(0.055) \\
   & BAND(0.3) & \textbf{0.789(0.089)} & 0.703(0.115) & 0.688(0.044) \\
   & BAND(0.6) & \textbf{0.806(0.104)} & 0.702(0.121) & 0.671(0.049) \\
   & CS(0.2) & \textbf{0.694(0.074)} & 0.599(0.070) & 0.648(0.037) \\
		\bottomrule
	\end{tabular}
\end{table}

\clearpage
\begin{table}
	\centering
	\caption{Simulation: identification of G-E interactions. In each cell, mean AUC (se). There are a total of 40 nonzero interactions, with coefficients $\sim uniform(0.4, 0.6)$. }
	\label{tab6}
	\begin{tabular}{ccccc}
		\toprule
		$\xi$ & cor & robust & nonrobust & quantile \\
		\midrule
		\multicolumn{5}{c}{continuous}\\
		\hline
		\multirow{6}{*}{0} & AR(0) & \textbf{0.647(0.048)} & 0.605(0.053) & 0.606(0.049) \\
   & AR(0.3) & \textbf{0.755(0.059)} & 0.707(0.079) & 0.752(0.058) \\
   & AR(0.8) & \textbf{0.892(0.074)} & 0.781(0.073) & 0.814(0.039) \\
   & BAND(0.3) & \textbf{0.791(0.074)} & 0.726(0.093) & 0.769(0.052) \\
   & BAND(0.6) & \textbf{0.838(0.066)} & 0.765(0.084) & 0.629(0.048) \\
   & CS(0.2) & \textbf{0.686(0.077)} & 0.609(0.055) & 0.602(0.048) \\
		\hline
		\multirow{6}{*}{0.1} & AR(0) & \textbf{0.623(0.041)} & 0.593(0.049) & 0.563(0.059) \\
   & AR(0.3) & \textbf{0.701(0.074)} & 0.625(0.085) & 0.602(0.055) \\
   & AR(0.8) & \textbf{0.863(0.060)} & 0.734(0.088) & 0.778(0.041) \\
   & BAND(0.3) & \textbf{0.750(0.070)} & 0.661(0.099) & 0.711(0.058) \\
   & BAND(0.6) & \textbf{0.795(0.069)} & 0.722(0.089) & 0.725(0.061) \\
   & CS(0.2) & \textbf{0.662(0.059)} & 0.586(0.052) & 0.596(0.062) \\
		\hline
		\multirow{6}{*}{0.3} & AR(0) & \textbf{0.581(0.050)} & 0.537(0.037) & 0.503(0.055) \\
   & AR(0.3) & \textbf{0.656(0.064)} & 0.570(0.051) & 0.596(0.049) \\
   & AR(0.8) & \textbf{0.807(0.086)} & 0.647(0.085) & 0.679(0.058) \\
   & BAND(0.3) & \textbf{0.677(0.077)} & 0.580(0.072) & 0.618(0.061) \\
   & BAND(0.6) & \textbf{0.718(0.082)} & 0.612(0.088) & 0.604(0.042) \\
   & CS(0.2) & \textbf{0.642(0.060)} & 0.550(0.055) & 0.571(0.046) \\
		\midrule
		\multicolumn{5}{c}{categorical}\\
		\hline
		\multirow{6}{*}{0} & AR(0) & \textbf{0.640(0.051)} & 0.604(0.052) & 0.611(0.052) \\
   & AR(0.3) & \textbf{0.743(0.067)} & 0.706(0.089) & 0.733(0.041) \\
   & AR(0.8) & \textbf{0.887(0.070)} & 0.779(0.070) & 0.806(0.034) \\
   & BAND(0.3) & \textbf{0.761(0.069)} & 0.716(0.089) & 0.751(0.039) \\
   & BAND(0.6) & \textbf{0.820(0.076)} & 0.781(0.087) & 0.623(0.059) \\
   & CS(0.2) & \textbf{0.688(0.071)} & 0.598(0.051) & 0.601(0.033) \\
		\hline
		\multirow{6}{*}{0.1} & AR(0) & \textbf{0.619(0.051)} & 0.565(0.050) & 0.548(0.047) \\
   & AR(0.3) & \textbf{0.706(0.068)} & 0.637(0.091) & 0.647(0.054) \\
   & AR(0.8) & \textbf{0.842(0.092)} & 0.728(0.089) & 0.751(0.061) \\
   & BAND(0.3) & \textbf{0.735(0.077)} & 0.667(0.092) & 0.726(0.041) \\
   & BAND(0.6) & \textbf{0.771(0.078)} & 0.691(0.102) & 0.736(0.034) \\
   & CS(0.2) & \textbf{0.658(0.061)} & 0.568(0.046) & 0.589(0.052) \\
		\hline
		\multirow{6}{*}{0.3} & AR(0) & \textbf{0.597(0.053)} & 0.540(0.043) & 0.540(0.062) \\
   & AR(0.3) & \textbf{0.663(0.070)} & 0.579(0.066) & 0.604(0.038) \\
   & AR(0.8) & \textbf{0.794(0.091)} & 0.658(0.094) & 0.704(0.058) \\
   & BAND(0.3) & \textbf{0.682(0.073)} & 0.606(0.080) & 0.626(0.046) \\
   & BAND(0.6) & \textbf{0.717(0.086)} & 0.615(0.084) & 0.614(0.047) \\
   & CS(0.2) & \textbf{0.646(0.060)} & 0.558(0.048) & 0.604(0.055) \\
		\bottomrule
	\end{tabular}
\end{table}

\clearpage
\begin{table}
	\centering
	\caption{
Analysis of the TCGA lung adenocarcinoma data using the nonrobust method. The identified interactions are denoted as ``gene * environmental variable". For the interactions, values in ``()" are the stability results.}
	\label{tab1}
	\begin{tabular}{lr}
		\toprule
		effect & estimate $\times$100\\
		\midrule		
        age & 0.868  \\
		gender & 6.683  \\
		smoking pack years & 0.041  \\
		smoking history & -20.163  \\
		SPATA33 & -7.060  \\
		DNAJC21 & 5.237  \\
		EIF4EBP1 & 8.736  \\
		FAM160B1 & -0.030  \\
		KIAA1586 & 4.018  \\
		LRRC37A4P & 3.040  \\
		ST6GALNAC1 & 5.989  \\
		TM2D2 & 10.110  \\
		TMEM192 & -5.785  \\
		TROVE2 & -3.019  \\
		WIPI2 & 5.296  \\
		SPATA33 * smoking pack years & -0.084(0.812) \\
		DNAJC21 * smoking history & 5.245(0.986) \\
		EIF4EBP1 * smoking pack years & -0.087(0.977) \\
		FAM160B1 * gender & 11.844(0.982) \\
		KIAA1586 * age & 0.107(0.954) \\
		LRRC37A4P * smoking pack years & -0.205(0.998) \\
		LRRC37A4P * smoking history & -6.799(0.998) \\
		ST6GALNAC1 * smoking pack years & -0.149(0.989) \\
		TM2D2 * smoking pack years & -0.176(0.998) \\
		TMEM192 * gender & 9.853(0.995) \\
		TROVE2 * gender & 7.349(0.929) \\
		WIPI2 * smoking history & 13.420(0.995) \\
		
		\bottomrule
	\end{tabular}
\end{table}

\clearpage
\begin{table}
	\centering
	\caption{Analysis of the TCGA lung adenocarcinoma data using the quantile method. The identified interactions are denoted as ``gene * environmental variable". For the interactions, values in ``()" are the stability results.}
	\label{tab2}
	\begin{tabular}{lr}
		\toprule
		effect & estimate $\times$100\\
		\midrule
age & 0.891  \\
ATP6V1C1 & 0.252  \\
C1ORF27 & 7.321  \\
SDE2 & 0.337  \\
CD46 & 0.584  \\
DNAJC21 & 1.272  \\
KLHL7 & 0.932  \\
PTK2 & 9.426  \\
PVT1 & 1.148  \\
RAB3GAP2 & 0.845  \\
TSPAN3 & 8.557  \\
TWISTNB & 0.872  \\
WDR26 & 1.265  \\
WIPI2 & 7.227  \\
YWHAZ & 1.883  \\
ATP6V1C1 * age & 0.0172(0.724) \\
C1ORF27 * age & 0.295(0.899) \\
SDE2 * age & 0.0153(0.758) \\
CD46 * age & 0.0344(0.862) \\
DNAJC21 * age & 0.327(0.791) \\
KLHL7 * age & 0.372(0.514) \\
PTK2 * age & 1.074(0.927) \\
PVT1 * age & 0.876(0.711) \\
RAB3GAP2 * age & 0.923(0.757) \\
TSPAN3 * age & 1.388(0.942) \\
TWISTNB * age & 0.915(0.812) \\
WDR26 * age & 1.279(0.798) \\
WIPI2 * age & 1.891(0.906) \\
YWHAZ * age & 1.596(0.796) \\		
		\bottomrule
	\end{tabular}
\end{table}

\end{document}